\documentclass[12pt]{article}
\usepackage[dvips]{graphicx}
\usepackage{epsfig}
\usepackage{amsmath,amssymb,amsbsy}
\usepackage{amsfonts}
\begin{document}
\thispagestyle{empty}
\begin{center}

{\Large\bf{The statistical model for parton distributions\footnote{Invited talk at XX International Workshop on "`Deep-Inelastic Scattering and related subjects"', University of Bonn, Germany, March 26-30 2012, presented by F. Buccella, to appear in the conference proceedings}}}

\vskip1.0cm
{\bf Claude Bourrely}
\vskip 0.2cm
Aix-Marseille Universit\'e,\\
D\'epartement de Physique, Facult\'e des Sciences de Luminy,\\
13288 Marseille, Cedex 09, France\\
\vskip 0.3cm
{\bf Franco Buccella}
\vskip 0.2cm
Dipartimento di Scienze Fisiche, Universit\`a di Napoli,\\
Via Cintia, I-80126, Napoli
and INFN, Sezione di Napoli, Italy
\vskip 0.3cm
{\bf Jacques Soffer}
\vskip 0.2cm
Physics Department, Temple University\\
Barton Hall, 1900 N, 13th Street\\
Philadelphia, PA 19122-6082, USA
\end{center}
\vskip 0.8cm
\begin{center}
{\bf Abstract}
\end{center}
The phenomenological motivations, the expressions and the comparison
with experiment of the parton distributions inspired by the quantum
statistics are described. The Fermi-Dirac expressions for the quarks
and their antiparticles automatically account for the correlation between
the shape and the first moments of the valence partons, as well as the
flavor and spin asymmetries of the sea. One is able to describe with a
small number of parameters both unpolarized and polarized structure 
functions.
\vskip 1.3cm
Let us first recall some of the basic ingredients for building up the parton
distribution functions (PDF) in the statistical approach, as oppose to the
standard polynomial type
parametrizations, based on Regge theory at low $x$ and counting rules at large
$x$.
The fermion distributions are given by the sum of two terms \cite{bbs1}, the
first one, 
a quasi Fermi-Dirac function and the second one, a flavor and helicity
independent diffractive
contribution equal for light quarks. So we have, at the input energy scale
$Q_0^2=4 \mbox{GeV}^2$,
\begin{equation}
xq^h(x,Q^2_0)=
\frac{AX^h_{0q}x^b}{\exp [(x-X^h_{0q})/\bar{x}]+1}+
\frac{\tilde{A}x^{\tilde{b}}}{\exp(x/\bar{x})+1}~,
\label{eq1}
\end{equation}
\begin{equation}
x\bar{q}^h(x,Q^2_0)=
\frac{{\bar A}(X^{-h}_{0q})^{-1}x^{2b}}{\exp [(x+X^{-h}_{0q})/\bar{x}]+1}+
\frac{\tilde{A}x^{\tilde{b}}}{\exp(x/\bar{x})+1}~.
\label{eq2}
\end{equation}
Notice the change of sign of the potentials
and helicity for the antiquarks.
The parameter $\bar{x}$ plays the role of a {\it universal temperature}
and $X^{\pm}_{0q}$ are the two {\it thermodynamical potentials} of the quark
$q$, with helicity $h=\pm$. It is important to remark that the diffractive
contribution 
occurs only in the unpolarized distributions $q(x)= q_{+}(x)+q_{-}(x)$ and it
is absent in the valence $q_v(x)= q(x) - \bar {q}(x)$ and in the helicity
distributions $\Delta q(x) = q_{+}(x)-q_{-}(x)$ (similarly for antiquarks).
The {\it eight} free parameters\footnote{$A=1.74938$ and $\bar{A}~=1.90801$
are
fixed by the following normalization conditions $u-\bar{u}=2$, $d-\bar{d}=1$.}
in Eqs.~(\ref{eq1},\ref{eq2}) were
determined at the input scale from the comparison with a selected set of
very precise unpolarized and polarized Deep Inelastic Scattering (DIS) data
\cite{bbs1}. They have the
following values
\begin{equation}
\bar{x}=0.09907,~ b=0.40962,~\tilde{b}=-0.25347,~\tilde{A}=0.08318,
\label{eq3}
\end{equation}
\begin{equation}
X^+_{0u}=0.46128,~X^-_{0u}=0.29766,~X^-_{0d}=0.30174,~X^+_{0d}=0.22775~.
\label{eq4}
\end{equation}
For the gluons we consider the black-body inspired expression
\begin{equation}
xG(x,Q^2_0)=
\frac{A_Gx^{b_G}}{\exp(x/\bar{x})-1}~,
\label{eq5}
\end{equation}
a quasi Bose-Einstein function, with $b_G=0.90$, the only free parameter
\footnote{In Ref.~\cite{bbs1} we were assuming that, for very small $x$,
$xG(x,Q^2_0)$ has the same behavior as $x\bar q(x,Q^2_0)$, so we took $b_G = 1
+ \tilde b$. However this choice leads to a too much rapid rise of the gluon
distribution, compared to its recent  determination from HERA data, which
requires $b_G=0.90$.}, since $A_G=20.53$ is determined by the momentum sum
rule.
 We also assume that, at the input energy scale, the polarized gluon 
distribution vanishes, so $x\Delta G(x,Q^2_0)=0$. For the strange quark
distributions, the simple choice made in Ref. \cite{bbs1}
was greatly improved in Ref. \cite{bbs2}. More recently, new tests against
experimental (unpolarized and
polarized) data turned out to be very satisfactory, in particular in hadronic
collisions, as reported in Refs.~\cite{bbs3,bbs4}.

For illustration, we will just give one recent result, directly related to the
determination of the quark distributions from unpolarized DIS.
 We display on Fig.~1 ($\it{Left}$), the resulting unpolarized statistical PDF
versus $x$ at $Q^2$=10 $\mbox{GeV}^2$, where $xu_v$ is the $u$-quark valence,
$xd_v$ the $d$-quark valence, with their characteristic maximum around $x=0.3$,
$xG$ the gluon and $xS$
stands for twice the total antiquark contributions, $\it i.e.$ $xS(x)=2x(\bar
{u}(x)+ \bar {d}(x) + \bar {s}(x))+ \bar {c}(x))$. Note that $xG$ and $xS$ are
downscaled by a factor 0.05. They can be compared with the parton distributions
as determined by the HERAPDF1.5 QCD NLO fit, shown also in
Fig.~1 ($\it{Right}$), and there is a good agreement. The results are based on
recent $ep$ collider data from HERA, combined with previously published data
and the accuracy is typically
in the range of 1.3 - 2$\%$.

\vspace*{-15mm}
\begin{figure}[htp]
\begin{center}
  \begin{minipage}{6.5cm}
\includegraphics[width=6.5cm]{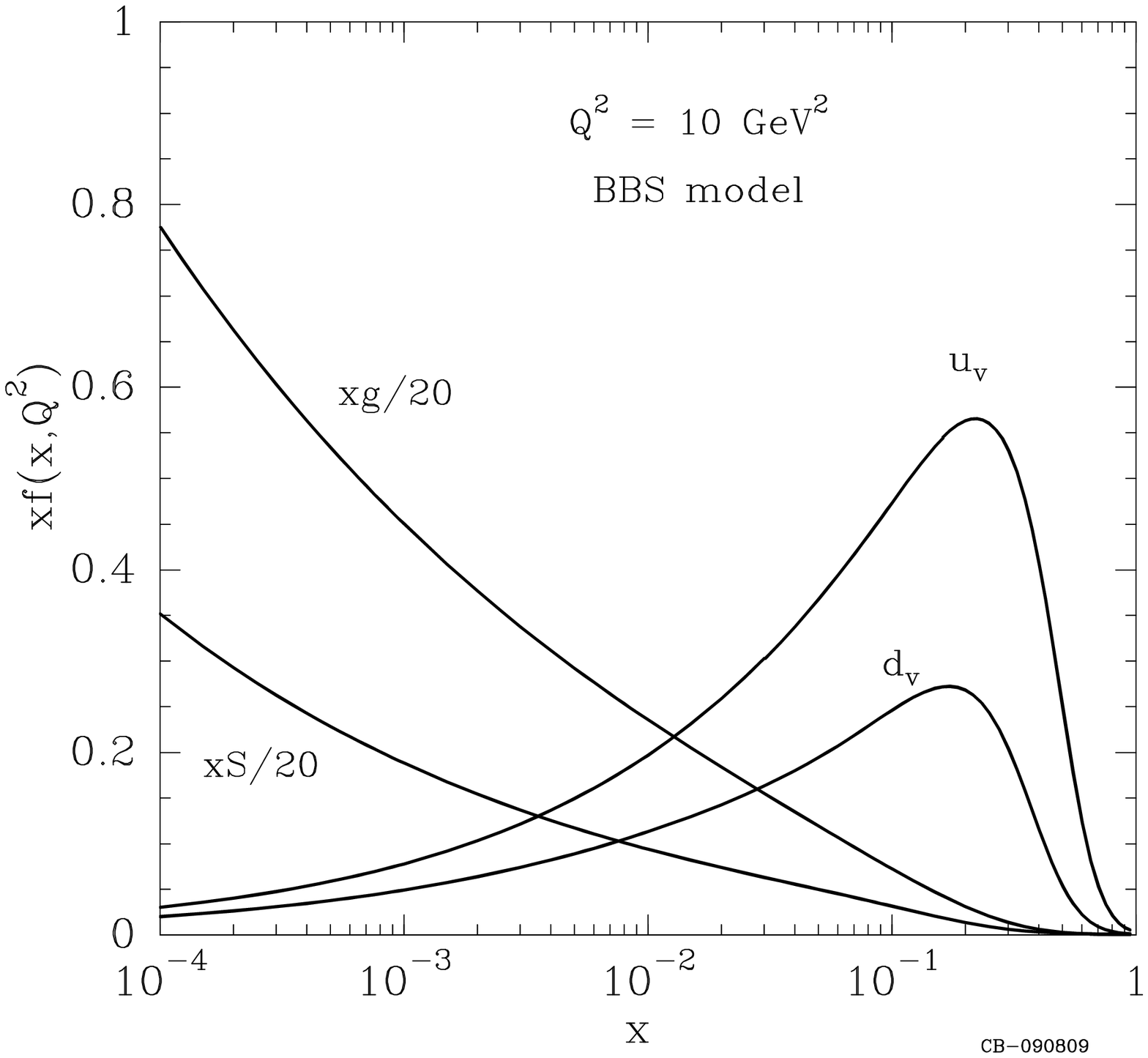}
  \end{minipage}
  \begin{minipage}{6.5cm}
\vspace*{10mm}
\includegraphics[width=7cm]{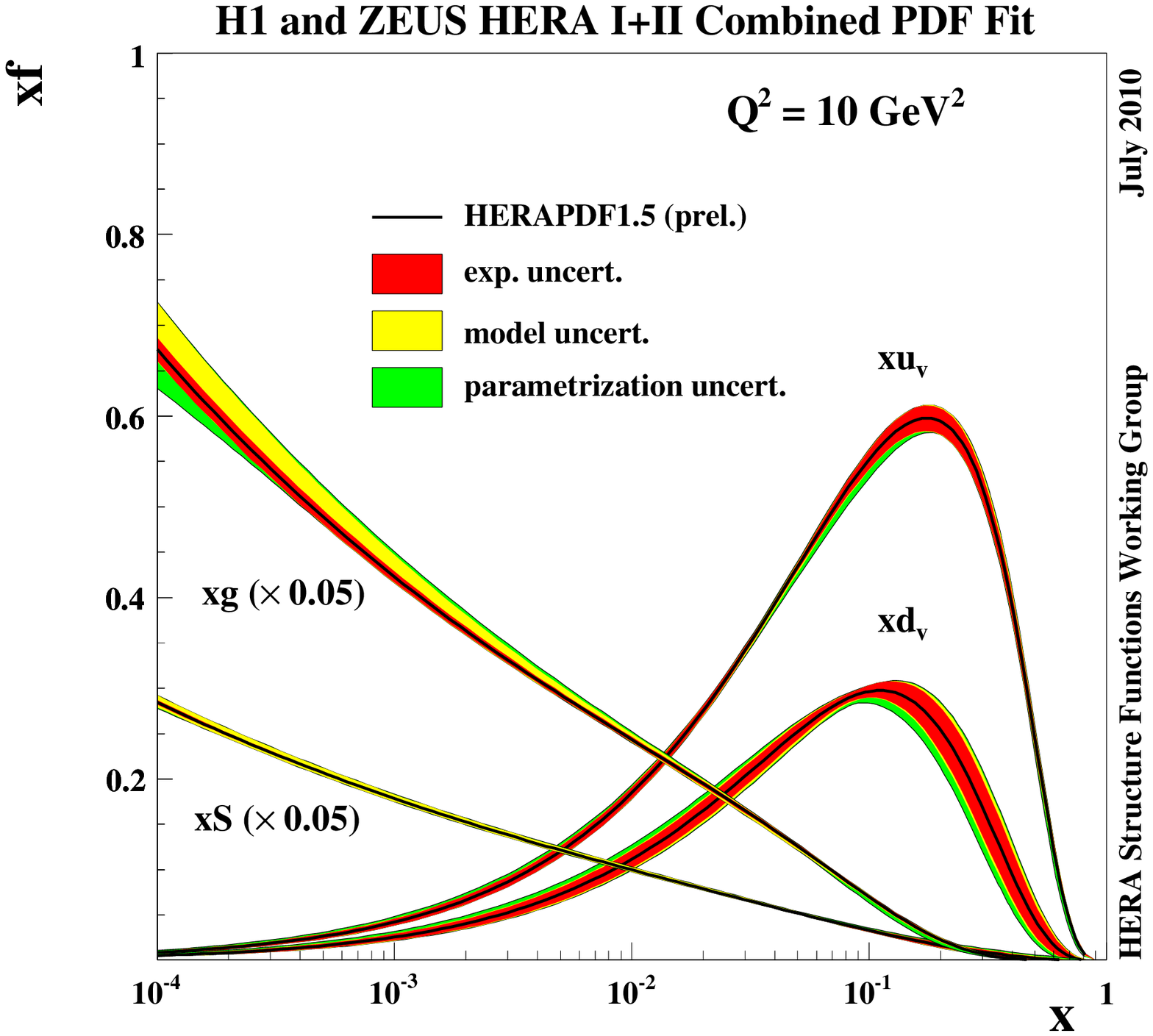}
  \end{minipage}
\end{center}
\vspace*{-8mm}
\caption{
{\it Left} : BBS predictions for various statistical unpolarized parton
distributions versus $x$ 
at $Q^2=10\mbox{GeV}^2$. {\it Right} : Parton distributions at
$Q^2=10\mbox{GeV}^2$, as determined 
by the HERAPDF fit, with different uncertainties (Taken from Ref.
\cite{aaron}).}
\label{fi:fig1}
\end{figure}

Another interesting 
point concerns the behavior of the ratio $d(x)/u(x)$, 
which depends on the mathematical properties of the ratio of two Fermi-Dirac 
factors, outside the region dominated by the diffractive contribution. 
So for $x>0.1$, this ratio is expected to decrease faster for 
$X_{0d}^+ - \bar x < x < X_{0u}^+ + \bar x$ and then above, for 
$x > 0.6$ it flattens out.
This change of slope is clearly 
\begin{figure}[htp]
\begin{center}
  \begin{minipage}{6.5cm}
\includegraphics[width=6.8cm,height=7.5cm]{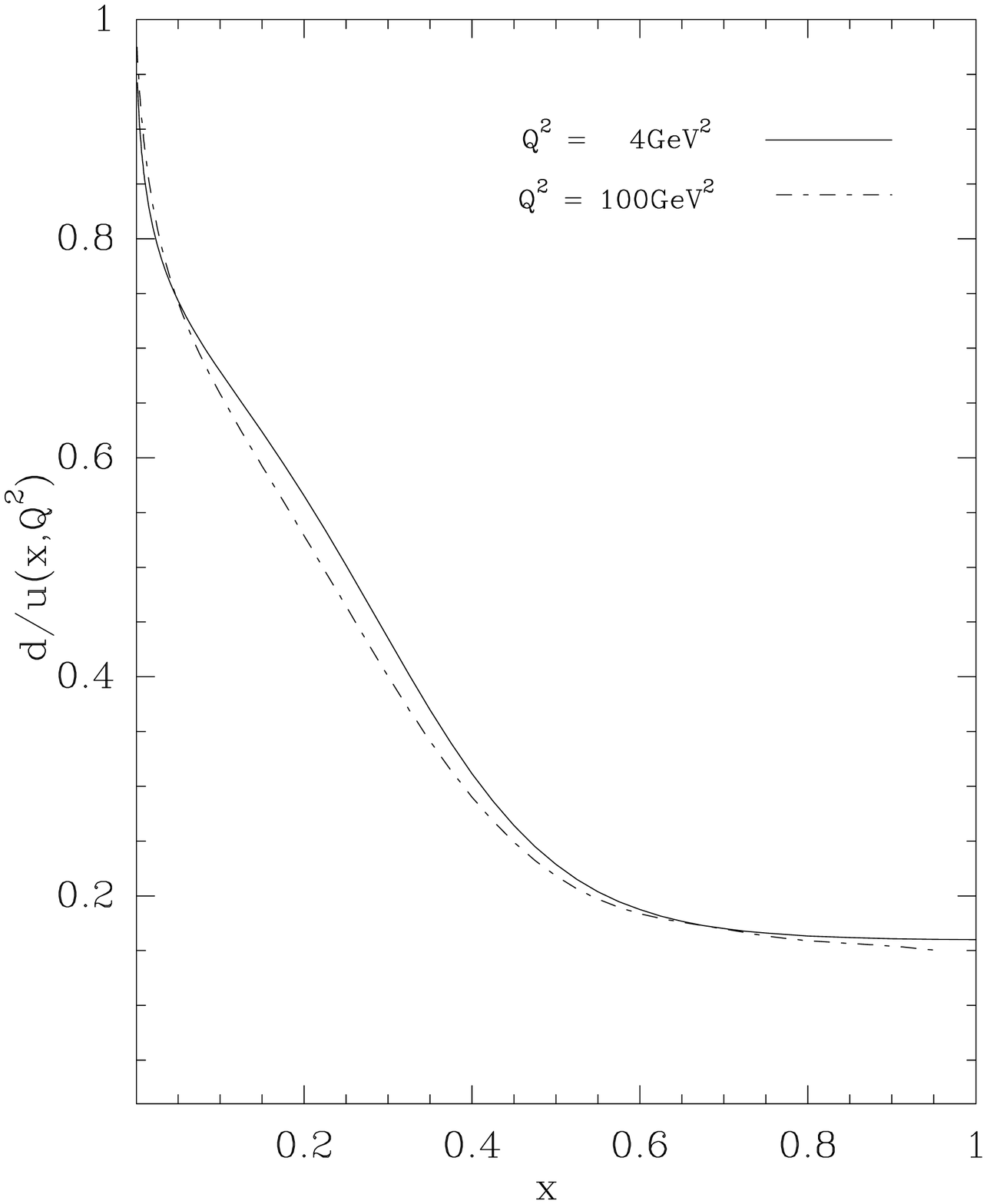}
  \end{minipage}
  \begin{minipage}{6.5cm}
\includegraphics[width=6.0cm,height=6.8cm]{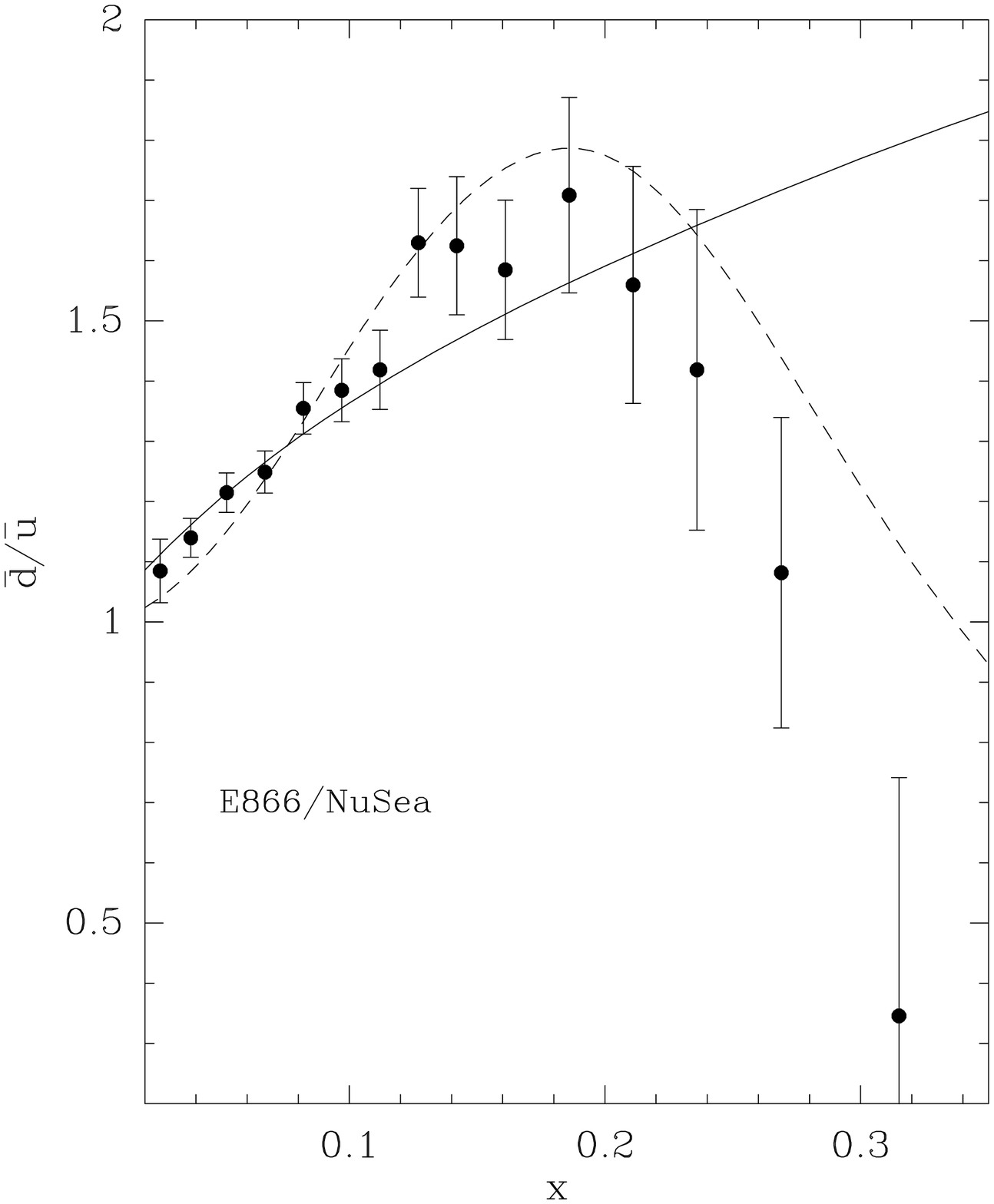}
  \end{minipage}
\end{center}
\vspace*{-5mm}
\caption{{\it Left} : The ratio $d(x)/u(x)$ as function of $x$ for $Q^2 =
4\mbox{GeV}^2$ 
(solid line) and $Q^2 =100\mbox{GeV}^2$ (dashed-dotted line). {\it Right} :
Comparison of the data on $\bar d / \bar u $ versus $x$, from E866/NuSea at
$Q^2=54\mbox{GeV}^2$
\cite{E866}, with the prediction of the statistical model (solid curve) 
and the set 1 of the parametrization proposed in Ref. \cite{Sassot}
(dashed curve).}
\label{fi:doveru}
\end{figure}
visible in Fig.~\ref{fi:doveru} ({\it Left}),
with a very 
little $Q^2$ dependence. Note that our prediction for the large $x$ behavior,
differs from most of the current literature, namely $d(x)/u(x) \to 0$
for $x \to 1$, but we find $d(x)/u(x) \to 0.16$ near the value $1/5$,
a prediction
originally formulated in Ref.~\cite{FJ}.
This is a very challenging question, since the very high-$x$ region remains
poorly known.\\
To continue our tests of the unpolarized parton distributions, we must come 
back to the important question of the flavor asymmetry of the light
antiquarks. Our determination of $\bar u(x,Q^2)$ and
$\bar d(x,Q^2)$ is perfectly consistent with the violation of the Gottfried
sum rule, for which we found the value $I_G= 0.2493$ for $Q^2=4\mbox{GeV}^2$.
Nevertheless there remains an open problem with the $x$ distribution
of the ratio $\bar d/\bar u$ for $x \geq 0.2$.
According to the Pauli principle, this ratio is expected to remain above 1 for
any value of
$x$. However, the E866/NuSea Collaboration \cite{E866} has
released the final results corresponding to the analysis of their full
data set of Drell-Yan yields from an 800 GeV/c proton beam on hydrogen
and deuterium targets and, for $Q^2=54\mbox{GeV}^2$, they obtain the ratio 
$\bar d/\bar u$ shown in Fig.~\ref{fi:doveru} ({\it Right}). 
Although the errors are rather large in the high-$x$ region,
the statistical approach disagrees with the trend of the data.
Clearly by increasing the number of free parameters, it
is possible to build up a scenario which leads to the drop off of
this ratio for $x\geq 0.2$.
For example this was achieved in Ref. \cite{Sassot}, as shown 
by the dashed curve in Fig.~\ref{fi:doveru} ({\it Right}). There is no such
freedom in the statistical
approach, since quark and antiquark distributions are strongly related. On the
experimental side, there are now new
opportunities for extending the $\bar d/ \bar u$ measurement to larger $x$ up
to $x=0.7$, 
with the running E906 experiment at the 120 GeV Main Injector at Fermilab
\cite{E906} and a proposed
experiment at the new 30-50 GeV proton accelerator at J-PARC \cite{JPARC}.

\begin{figure}[htp]
\vspace*{-15mm}
\begin{center}
  \begin{minipage}{6.5cm}
\includegraphics[width=7.0cm]{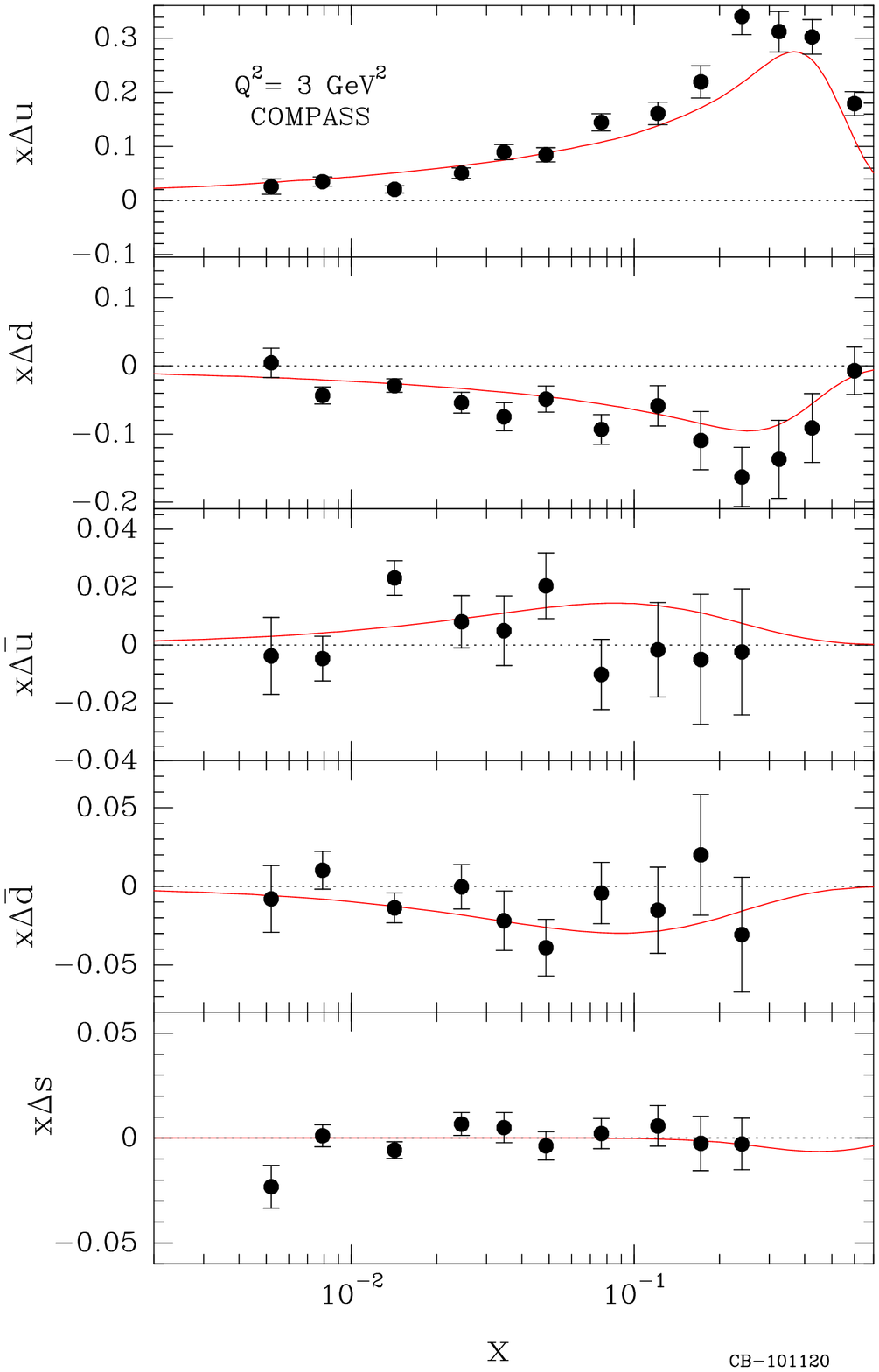}
  \end{minipage}
  \begin{minipage}{6.5cm}
\vspace*{10mm}
\includegraphics[width=7.0cm,height=10.2cm]{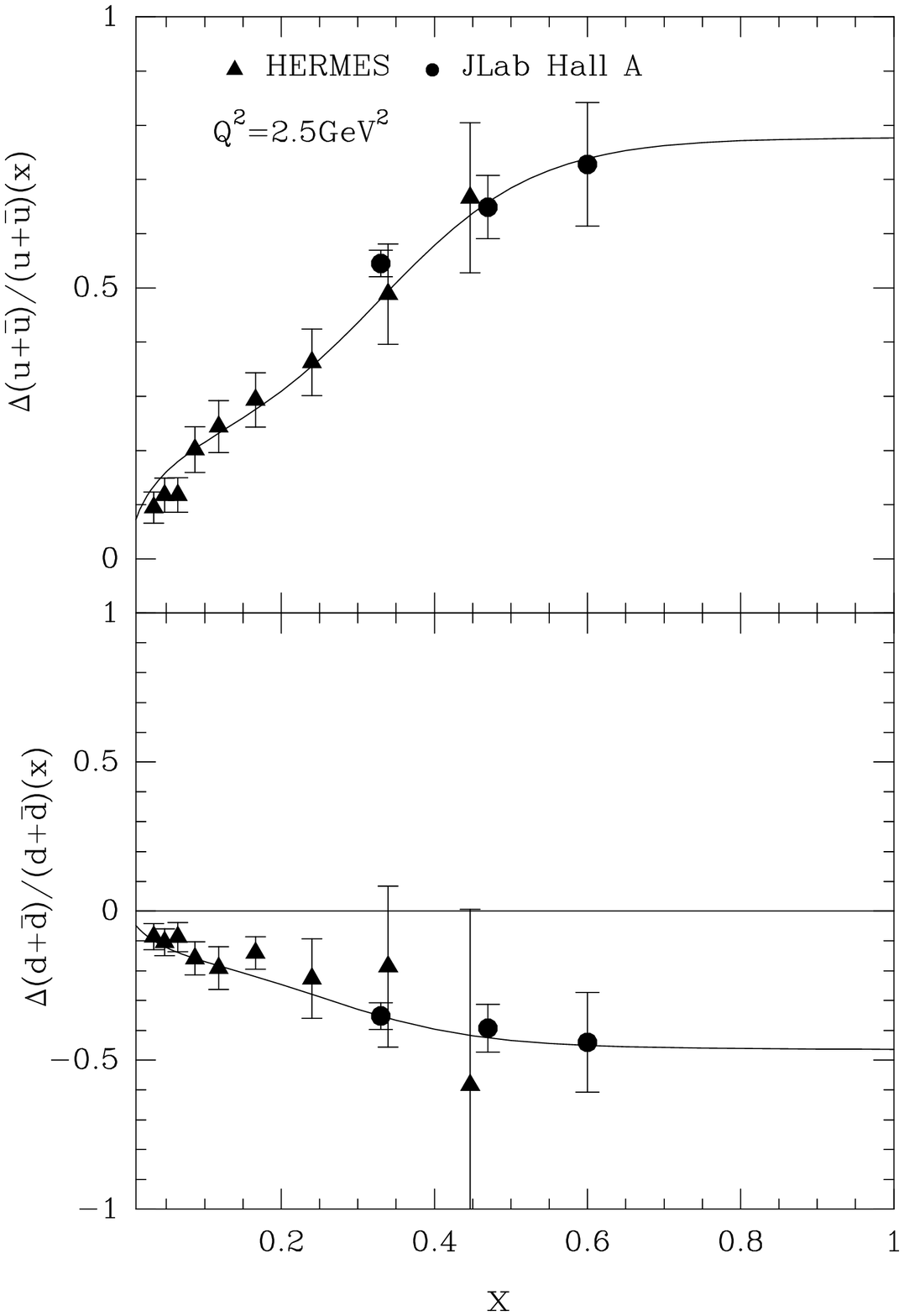}
  \end{minipage}
\end{center}
  \vspace*{-15mm}
\caption{{\it Left} : Quark and antiquark helicity distributions as a function
of $x$ for $Q^2 = 3\mbox{GeV}^2$. Data from 
COMPASS \cite{COMPASS}. The curves are
predictions from the statistical approach. {\it Right} : Ratios $(\Delta u +
\Delta \bar u)/(u + \bar u)$ and 
$(\Delta d + \Delta \bar d)/(d + \bar d)$ as a function of $x$.
Data from Hermes for $Q^2 = 2.5\mbox{GeV}^2$ \cite{herm99} and
a JLab Hall A experiment \cite{JLab04}. The curves are predictions from the
statistical approach.}
\label{fi:fig4}
\end{figure}
Analogous considerations can be made for the corresponding helicity 
distributions, whose most recent determinations are shown in Fig.~\ref{fi:fig4}
({\it Left}).
By using a similar argument as above, the ratio $\Delta u(x)/u(x)$ 
is predicted to have a rather fast increase in the $x$ range 
$(X^-_{0u}-\bar{x},X^+_{0u}+\bar{x})$
and a smoother behaviour above, while $\Delta d(x)/d(x)$, which is negative,
has a fast decrease in the $x$ range $(X^+_{0d}-\bar{x},X^-_{0d}+\bar{x})$ 
and a smooth one above. This is exactly the trends displayed in 
Fig.~\ref{fi:fig4} ({\it Right})  and our predictions are in perfect agreement
with the accurate high-$x$ data. We note the behavior near $x=1$, another
typical property of the statistical
approach, is also at variance with predictions of the current literature. 
The fact that $\Delta u(x)$ is more concentrated in the higher $x$ region than
$\Delta d(x)$, accounts for the change of sign of $g^n_1(x)$, which becomes
positive for $x>0.5$, as first observed at Jefferson Lab \cite{JLab04}.

Concerning the light antiquark helicity distributions, the statistical 
approach
imposes a strong relationship to the corresponding quark helicity
distributions. In particular, it predicts $\Delta \bar u(x)>0$ and $\Delta
\bar
d(x)<0$, with almost the same magnitude, in contrast with the
simplifying assumption $\Delta \bar u(x)=\Delta \bar d(x)$, often adopted in
the literature.
The COMPASS data \cite{compass} give $\Delta \bar u(x) + \Delta \bar d(x)
\simeq 0$, which implies either small or
opposite values for $\Delta \bar u(x)$ and $\Delta \bar d(x)$. Indeed $\Delta
\bar u(x)>0$ and $\Delta \bar d(x)<0$, predicted by
the statistical approach \cite{bbs1} (see Fig.~\ref{fi:fig4} ({\it Left}), lead
to a non negligible
 positive contribution of the sea to the Bjorken sum rule, an interesting
consequence.
 For lack of space we only mention the extension to the transverse momentum
dependence (TMD), an important aspect of the statistical PDF and we refer the
reader to Ref. \cite{bbs6}.

A new set of PDF was constructed in the framework of a statistical
approach of the nucleon.
All unpolarized and polarized distributions depend upon {\it nine}
free parameters for light quarks and gluon, with some physical meaning.
 New tests against experimental (unpolarized and polarized)
data on DIS, Semi-inclusive DIS and also hadronic processes, are very
satisfactory.
It has a good predictive power, but some special features remain to be
verified, specially in the high-$x$ region, a serious challenge for the
future.

\end{document}